\documentstyle[prb,aps,preprint]{revtex}
\input {psfig}

\newcommand{\bv}[1]{\mbox{\boldmath$ #1$}} 

\begin{document}
\draft
\title{Nonlinearity in NS transport: scattering matrix approach}

\author{G.~B.\ Lesovik$^{a}$, A.~L.\ Fauch\`ere$^{b}$, G.\ Blatter$^{b}$}

\address{$^{a}$Institute of Solid State Physics, Chernogolovka 142432,
 Moscow district, Russia\\
$^{b}$Theoretische Physik, Eidgen\"ossische Technische Hochschule,
 CH-8093 Z\"urich, Switzerland }

\maketitle

\begin{abstract}
  A general formula for the current through a disordered
  normal--superconducting junction is derived, which is valid at
  finite temperature and includes the full voltage dependence.
  The result depends on a multichannel scattering matrix,
  which describes elastic scattering in the normal region,
  and accounts for the Andreev scattering at the NS interface.
  The symmetry of the current with respect to sign reversal in the
  subgap regime is discussed. The Andreev approximation is used to
  derive a spectral conductance formula, which applies to voltages
  both below and above the gap. In a case study the spectral
  conductance formula is applied to the problem of an NINIS double 
  barrier junction. 
\end{abstract}

\section{Introduction}

The study of electronic transport in normal-metal -- superconductor (NS)
or semiconductor -- superconductor (SmS) sandwiches has attracted a
considerable amount of interest in the past years. At sufficiently low
temperature and in high quality mesoscopic samples, the phase-breaking
length of the electrons is larger than the typical system size.  The
resulting coherence of the electron wavefunctions produces directly
observable quantum effects, which manifest themselves in a typical
non-local response of the currents to the applied fields. Of special
interest is the effect of the electronic phase-coherence in a normal 
metal -- superconductor system. In the standard theory of the proximity 
effect, the influence of the superconductor on the normal metal can be 
understood in terms of the coherent coupling of electrons and holes in 
the metal as described by the Bogoliubov-de Gennes (BdG) equations
\cite{deGennes}. The correlation between
electrons and holes is produced by the process of Andreev
reflection\cite{andreev,btk} at the NS interface, which mixes electron
and hole states while quasiparticle current is converted to
supercurrent. This microscopic picture of electron-hole correlation is 
equivalent to that provided by a condensate wavefunction in the normal metal 
which is induced by the superconductor through the continuity conditions at the
NS interface. The scattering matrix approach makes use of the microscopic 
single particle picture of coupled electron and hole channels providing a
straightforward and powerful tool within a formalism of the 
Landauer--B\"uttiker type\cite{beenakker,landauer,buttland}. 

The quality of the interface as well as the phase breaking processes 
determine the strength of the proximity effect and naturally have their 
impact on the current--voltage characteristics (CVC). A few fascinating 
transport experiments\cite{pothier,petrashov,pannetier} have been carried 
out recently, investigating temperature and voltage dependence, as well as
the flux modulation, of both NS and SmS junctions. Interestingly, the
relative strength of the interface barrier and the elastic scattering
in the normal region is crucial for the features of the CVC. The ratio
of the two determines whether subgap conductance peaks arise at zero or 
finite voltage\cite{volkov,marmokos,yip}.  The investigation of these 
so-called zero and finite bias anomalies in the subgap conductance have 
been the object of recent experiments
\cite{kastalsky,nguyen,nitta,bakker,magnee,sanquer}. 
The present work draws much of its motivation from the 
ongoing discussions and experiments in this area. In our case study of a
double barrier NINIS junction we observe zero and finite bias anomalies and 
shed light on the mechanism producing these structures.

Stoof and Nazarov\cite{nazarov} have recently described carrier transport 
in disordered NS junctions in terms of an energy-dependent diffusion 
constant, successfully explaining the recent experiments on reentrance in the
conductivity at low temperatures\cite{petrashov,pannetier}. 
Their work uses the quasi-classical Green's 
functions technique, which allows to describe transport both close to 
equilibrium\cite{hekking} and far away from equilibrium\cite{volkov,nazarov}
 and facilitates the averaging over disorder in diffusive conductors. 
This approach is quite indispensable if phase-breaking processes are to be 
included. An appealing alternative approach is the scattering matrix technique, 
which relies on the quasiparticle wavefunctions described by the BdG equations. 
While being valid in a general context, it describes the transport in
mesocopic systems from a ballistic point of view, suggesting an
intuitive, nearly classical picture\cite{landauer,buttland}. The transport
through normal or superconducting leads is expressed through the properties of
a multichannel scattering matrix accounting for all elastic scattering 
processes, whether they be due to a geometric constriction, single impurities, 
or disorder (inelastic processes are excluded from such a description). 
By these means, the transport problem is reduced to solving a ballistic problem 
at the interfaces of the normal and superconducting leads. The current and the
conductance of the system are determined analytically in terms of the 
transmission and reflection amplitudes of the scattering matrix. 
Adhering to this formalism rather than the Green's function technique 
helps us to improve our understanding of zero and finite bias anomalies.

The study of normal--superconducting junctions goes back to the works
of Kulik\cite{kulik} on SNS junctions and of Blonder et al.\cite{btk}
 on NIS junctions, who studied nonlinear transport within the framework 
of the BdG equations considering quasi one-dimensional models. On the other
hand, the scattering matrix technique was developed by Landauer and 
co-workers\cite{landauer,buttland} in the linear response regime,
resulting in the well known conductance formula for a normal metal. 
Lambert\cite{lambert} and Takane and Ebisawa
\cite{takane} extended the approach to include superconducting segments, on
the basis of which Beenakker\cite{beenakker} derived a zero temperature, 
linear response conductance formula for the transport through NS junctions.
A few studies have been carried out recently\cite{marmokos,brouwer} combining
 the scattering matrix approach with the finite voltage transport model of 
Ref.\cite{btk}. They have limited themselves to the energy 
dependence of the scattering states to extract finite voltage properties of
the CVC in the subgap regime.
Here, we extend these works to voltages above the gap and additionally 
take into account the full voltage dependence of the transport problem. 
This is important within the context of the sign reversal symmetry of the
differential conductance which we discuss below. At the same time, we provide 
a common framework for the above studies, tracing them back to a single 
general formula.

In the present paper, we derive the general expression for the current through 
a NS junction in the scattering matrix approach, valid for multiple channels, 
finite voltage, and nonzero temperature. In section II, we present the 
derivation of the current--voltage relation and express it in terms of a 
{\it spectral} conductance formula, thereby accounting for the full voltage
dependence of the transport problem. We discuss the related symmetry of the CVC 
with respect to the sign reversal of the bias, which follows independently of 
any approximations for the BdG equations. In section III, the reflection at 
the NS interface is made explicit using the Andreev approximation and 
a spectral conductance formula is obtained expressing the result in terms of
the normal, energy and voltage dependent scattering matrix. We illustrate this
formula in the new regime of voltages above the gap and expose its 
connection to previously obtained limits. In section IV, the 
conductance formula is used for the analytical as well as the 
numerical analysis of a multichannel, double barrier NS junction. 
We describe the existence of resonances due to 
quasi-bound Andreev states, and show that they produce sharp conductance
peaks in both the single and multichannel junction. We explain the 
generic mechanism underlying the appearance of zero and finite bias 
anomalies in the ballistic two barrier system. Furthermore, we interpret
our results in connection to experiments and theories of zero and finite
bias anomalies in disordered NS junctions.

\section{Current voltage relation and spectral conductance}

We consider a normal--superconducting junction with quasi
one-dimensional, ballistic normal and superconducting leads, as shown
in Figure 1. The pair potential vanishes in the normal part, due to
the absence of attractive electron-electron interactions. Between the
normal lead and the NS interface, the electrons traverse a disordered
region, the transmission through and the reflection thereof are
described by a scattering matrix. The disorder stands for any source
of elastic scattering processes, including effects of a magnetic
field. Between the scattering region and the NS interface, a small
ballistic normal region serves to separate the scattering in the
normal part, which mixes all electron channels at a given energy,
from the scattering at the NS interface, where electron and hole
channels are mixed in the reflection process (the evanescent modes in
this intermediate ballistic region are neglected). The coherent
scattering in the disorder region of the normal metal is described by
the $4N\times 4N$ scattering matrix,
\begin{equation} \small \left(\begin{array}{c}O_1^e\\ O_1^h\\O_2^e\\
 O_2^h \end{array}\right)=\left(\begin{array}{cccc}
r_{11}\left(\epsilon\right) & 0 & t_{12}\left(\epsilon\right) & 0 \\ 
0 & r_{11}^*\left(-\epsilon\right) & 0 & t_{12}^*\left(-\epsilon\right)
 \\ t_{21}\left(\epsilon\right) & 0 & r_{22}\left(\epsilon\right) & 0 \\ 
0 & t_{21}^*\left(-\epsilon\right) & 0 & r_{22}^*\left(-\epsilon\right) 
\end{array}\right) \left(\begin{array}{c}I_1^e\\ I_1^h\\ I_2^e\\ I_2^h 
\end{array}\right) = \Large \left( \begin{array}{cc}\hat{r}_{11}
\left(\epsilon\right) & \hat{t}_{12}\left(\epsilon\right)\\ 
\hat{t}_{21}\left(\epsilon\right) & \hat{r}_{22}\left(\epsilon\right) 
\end{array}\right) \small \left(\begin{array}{c}I_1^e\\ I_1^h\\ I_2^e\\ 
I_2^h \end{array}\right), \label{scattmat} \end{equation}
which we denote by $\bv{S}_N$. The matrix connects the incoming $N$
electron (hole) channels $I_i^e$ ($I_i^h$) on each side to the equal
energy outgoing channels $O_i^e$ ($O_i^h$) according to Figure 1
($i=1,2$). The $N$ channels represent the different transverse states
at the Fermi surface (we neglect a change of the number of channels
with increasing voltage, which is strictly valid only in the presence
of electron-hole symmetry\cite{elhole}). $r_{ii}$ and $t_{ij}$ are
$N\times N$ reflection and transmission matrices for electron
channels, $\hat{r}_{ii}$ and $\hat{t}_{ij}$ the comprehensive
$2N\times 2N$ matrices including the complex conjugated reflection and
transmission amplitudes for holes. (Here we will denote the complex
conjugate of a matrix $m$ by $m^*$, the transposed matrix by
$m^{\top}$ and the adjoint by $m^{\dag}$.) Following usual convention, 
we include the propagation in the ballistic region $N_2$ in the
scattering matrix. For states normalized to carry unit probability
current\cite{prob}, the continuity equation implies the unitarity of
the scattering matrix. Note that the important distinction to previous
work is, that we allow the scattering matrix not only to depend on the
energy of the states, but also intrinsically on the applied voltage (the
voltage dependence is not explicit in the notation). This is necessary to
account for the full voltage dependence of the scattering problem. We can
thus describe the deformation of the states due to the space dependent 
potential including, e.g., the voltage dependent Schottky barrier at a 
semiconductor--superconductor (SmS) interface.

We define an analogous unitary scattering matrix for the NS interface 
$\bv{S}_I$ by
\begin{equation} \small \left(\begin{array}{c}I_2^e\\ I_2^h\\O_S^e\\ 
O_S^h \end{array}\right)=\left(\begin{array}{cccc}r_{ee}\left(\epsilon
\right) & r_{eh}\left(\epsilon\right) & t'_{ee}\left(\epsilon\right) & 
t'_{eh}\left(\epsilon\right)\\ r_{he}\left(\epsilon\right) & r_{hh}
\left(\epsilon\right) & t'_{he}\left(\epsilon\right) & t'_{hh}\left(
\epsilon\right)\\ t_{ee}\left(\epsilon\right) & t_{eh}\left(\epsilon
\right) & r'_{ee}\left(\epsilon\right) & r'_{eh}\left(\epsilon\right)\\ 
t_{he}\left(\epsilon\right) & t_{hh}\left(\epsilon\right) & r'_{he}
\left(\epsilon\right) & r'_{hh}\left(\epsilon\right) \end{array}\right) 
\left(\begin{array}{c}O_2^e\\ O_2^h\\ I_S^e\\ I_S^h \end{array}\right) 
= \Large \left(\begin{array}{cc}\hat{r}_{I}\left(\epsilon\right) & 
\hat{t}'_{I}\left(\epsilon\right)\\  \hat{t}_{I}\left(\epsilon\right) & 
\hat{r}'_{I}\left(\epsilon\right)  \end{array}\right) \small \left(
\begin{array}{c}O_2^e\\ O_2^h\\ I_S^e\\ I_S^h \end{array}\right). 
\label{ascattmat} \end{equation}
Incident and outgoing channels are again labeled according to Figure
1. $r$, $r'$ and $t$, $t'$ are $N\times N$ reflection and transmission
matrices between states normalized to unit probability current in the
normal and superconducting lead\cite{probcharge}, and are grouped into the 
$2N\times 2N$ matrices $\hat{r}_I$, $\hat{r}'_I$, $\hat{t}_I$, 
$\hat{t}'_I$.  In the evaluation of the current, we will consider the 
matrix (\ref{scattmat}) to be specified by an arbitrary model 
for the disorder, while the matrix (\ref{ascattmat}) will be made explicit 
below using the BdG equations. For the
present purpose they are both arbitrary and thus the shape of the
pair potential need not be specified. The total effect of all
scattering processes in the disorder region and at the NS interface can be
described by a global scattering matrix of the type (\ref{ascattmat}),
which is unitary as well. We restrict ourselves to its submatrix
$\hat{R}$ describing the reflection into the normal region,
\begin{equation} \small \left(\begin{array}{c}O_1^e\\ O_1^h \end{array}
\right)= \Large \hat{R}\left(\epsilon, V\right) \small \left(\begin{array}
{c}I_1^e\\ I_1^h \end{array}\right) = \left( \begin{array}{cc}R_{ee}
\left(\epsilon,V\right) & R_{eh}\left(\epsilon,V\right)\\ R_{he}\left(
\epsilon,V\right) & R_{hh}\left(\epsilon,V\right) \end{array} \right) 
\left(\begin{array}{c}I_1^e\\ I_1^h \end{array}\right) \label{globalR}. 
\end{equation}
$R_{ee}$, $R_{eh}$, $R_{he}$, and $R_{hh}$ are again $N\times N$
reflection matrices. $\hat{R}\left(\epsilon,V\right)$ will be computed
below from the given scattering matrices (\ref{scattmat}) and
(\ref{ascattmat}). We now derive the current--voltage relation based on
this global reflection matrix (we do not need the transmission matrix, 
since we determine the currents in the normal lead). 

Applying a voltage $V$ (denoting the voltage in a two point measurement) on
the normal side has two consequences. First, the voltage induces an 
electrostatic potential drop over the disorder region in the NS junction, 
resulting in a deformation of scattering states. The coupling of incident 
and outgoing channels is thus voltage dependent in general, as described by
$\hat{R}\left(\epsilon, V\right)$. A stationary state
incident from the normal lead (of energy $\epsilon$ in channel $\nu$)
consists of the incident electron and the reflected electron- and hole-
states and carries the current ($e=|e|$),
\begin{equation} I_{\nu}\left(\epsilon, V\right)=\frac{-e\hbar k_{\nu}}{m}
  \left\{1-\sum_{\beta} \mid
    R_{ee}\left(\epsilon,V\right)_{\beta\nu}\mid^2 + \sum_{\beta} \mid
    R_{he}\left(\epsilon,V\right)_{\beta\nu}\mid^2 \right\}.
\end{equation} 
Second, the applied voltage shifts the chemical
potential of the reservoir attached to the normal lead by $-eV$ with
respect to the reservoir on the superconducting side. The deformation
of the states by itself produces no net current\cite{net}.
The net current flow results exclusively
from the difference in occupation of the (finite voltage) scattering
states incident from the left and right reservoirs. Writing the sum over 
channels as a trace, we obtain the current--voltage relation
\begin{equation} I= \int d\epsilon\, \frac{1}{e}\Big[f\left(\epsilon\right)
-f\left(\epsilon+eV\right)\Big]\, G_s\left(\epsilon,V\right), 
\label{current} \end{equation}
with the spectral conductance
\begin{equation} G_s\left(\epsilon,V\right)=\frac{2e^2}{h}\mbox{Tr}
\bigg\{1-R_{ee}^{\dag}\left(\epsilon,V\right) R_{ee}\left(\epsilon,V
\right)+R_{he}^{\dag}\left(\epsilon,V\right) R_{he}\left(\epsilon,V
\right)\bigg\}. \label{cond} \end{equation}
A factor two accounts for the spin degeneracy of the channels. The defined
{\it spectral} conductance $G_s\left(\epsilon,V\right)$ describes the 
current contribution of the incident scattering states at energy 
$\epsilon$, at a given voltage $V$  (by convention, 
the energy is measured with respect to the chemical potential in the
superconductor). Formulas (\ref{current}) and (\ref{cond}) imply the 
differential conductance
\begin{eqnarray} \frac{dI}{dV}\mid_V &=& -\int d\epsilon\, f'\left(\epsilon
+eV\right)\, G_s\left(\epsilon,V\right) \nonumber \\
&& +\int d\epsilon\, \frac{1}{e}\Big[f\left(\epsilon\right)-f\left(\epsilon
+eV\right)\Big]\, \frac{\partial G_s\left(\epsilon,V\right)}{\partial V},
\end{eqnarray}
with the expansion
\begin{equation} \frac{dI}{dV}\mid_V = G_s\left(-eV,0\right) + 2V \partial_V 
G_s\left(\epsilon,V\right)\mid_{\epsilon=-eV,V=0} + \mbox{... }, \end{equation}
at zero temperature.
This differs from the differential conductance of Ref.~\cite{btk}, 
$dI/dV=G_s\left(-eV,0\right)$, by accounting for the change in the conductance
of the open channels with increasing voltage.

To complete the general derivation we need to express the matrix
$\hat{R}$ in terms of the given scattering matrices (\ref{scattmat}) and
(\ref{ascattmat}). Summing over all scattering paths of an incident 
electron or hole excitation, multiply scattered between the disorder 
region and the NS interface, we arrive at
\begin{equation} \hat{R}\left(\epsilon, V\right) =\hat{r}_{11}\left(
\epsilon\right)+\hat{t}_{12}\left(\epsilon\right)  \Big[1-\hat{r}_{I}
\left(\epsilon\right)\hat{r}_{22}\left(\epsilon\right)\Big]^{-1}  
\hat{r}_{I}\left(\epsilon\right) \hat{t}_{21}\left(\epsilon\right). 
\label{Rhat} \end{equation}
Apart form the direct reflection at the disorder region, the simplest 
process contributing consists of an excitation, which is first
transmitted ($\hat{t}_{21}$) through the disorder region,
reflected ($\hat{r}_{I}$) at the NS interface, and transmitted 
($\hat{t}_{12}$) back to the normal lead. All further paths result form
iterative scattering processes between the disorder region and the NS
interface (note that the scattering matrix for the disorder includes
the propagation in the ballistic lead $N_2$). 
The expressions (\ref{current}), (\ref{cond}), and (\ref{Rhat})
determine the general form of the current--voltage relation of a
disordered NS junction, without having made any assumptions about the
nature of the scattering at the NS interface like, e.~g., the shape of
the pair potential $\Delta$. In the next section, we evaluate the spectral 
conductance (\ref{cond}) further by using the Andreev approximation for the
scattering at the NS interface.

We close this section with a discussion of the symmetry of the CVC with 
respect to sign reversal of the applied voltage. In the subgap regime 
$e|V|<\Delta$, the incoming quasiparticle excitations may not enter the 
superconductor. The probability current of the states $|\epsilon|<\Delta$
is totally reflected and thus the global reflection matrix
$\hat{R}\left(\epsilon,V\right)$ of (\ref{globalR}) is unitary. The
unitarity produces the relations $R_{ee}^{\dag}R_{ee} + R_{he}^{\dag}R_{he}=1$
and $R_{ee} R_{ee}^{\dag} + R_{eh} R_{eh}^{\dag}=1$.
The symmetry of electron- and hole- type excitations in the BdG equations 
guarantees $R_{eh}\left(\epsilon,V\right)=-R_{he}^{*}\left(-\epsilon,V\right)$. 
As a consequence, the subgap conductance takes the form
\begin{eqnarray} G_s\left(\epsilon,V\right)&=&\frac{4e^2}{h}\mbox{Tr}
\bigg\{R_{he}^{\dag}\left(\epsilon,V\right) R_{he}\left(\epsilon,V
\right)\bigg\} \nonumber = \frac{4e^2}{h}\mbox{Tr}
\bigg\{R_{eh}^{\dag}\left(\epsilon,V\right) R_{eh}\left(\epsilon,V
\right)\bigg\} \nonumber \\
&=& \frac{4e^2}{h}\mbox{Tr}
\bigg\{R_{he}^{\dag}\left(-\epsilon,V\right) R_{he}\left(-\epsilon,V
\right)\bigg\} = G_s\left(-\epsilon,V\right). \end{eqnarray}
A subtle issue is that this symmetry does not yet imply a symmetry in 
the CVC under reversal of voltage\cite{lamblead}. The latter requires
that $G_s\left(\epsilon,V\right)=G_s\left(-\epsilon,-V\right)$, which
amounts to $G_s\left(\epsilon,V\right)$ being independent of voltage. 
Then we have $G_s\left(\epsilon\right)|_{\epsilon=-eV}=dI/dV|_V$ and the 
differential conductance is invariant under sign reversal of the voltage. 
Indeed, in recent experiments on SmS junctions\cite{magnee,sanquer}, 
an asymmetry in the CVC was found in the subgap regime, which can be 
understood on the basis of the above discussion taking into account the
voltage dependent Schottky barrier at the SmS interface. 
The deviations from the symmetry are of the
order of $eV/\mu$ or $eV/V_{o}$, where $\mu$ denotes the chemical potential
and $V_{o}$ characterizes the strength of the scattering potential. 
An explicit account of the voltage dependence of $G_s$ requires the
scattering matrix $\bv{S}_{N}$ to be determined in the applied electrostatic
potential. In principle, this task demands the self-consistent solution of the
scattering problem and the Poisson equation\cite{buttiker}, a problem
which has not yet been thoroughly addressed. In several cases of interest, 
though, an approximate consideration of the voltage dependence will furnish
an  accurate description. 

\section{Spectral conductance in the Andreev approximation}

We start from the spectral conductance formula (\ref{cond}) and
evaluate it by solving the boundary conditions for the transparent
interface in the Andreev approximation. The stationary states in the
ballistic leads are solutions of the BdG equations\cite{deGennes} and
are of the plane wave type. A step function model for the pair
potential $\Delta\left(x\right)= \Delta_o e^{i\chi}\theta\left(x\right)$ 
is assumed, which neglects the suppression of the pair potential in the 
superconductor on the distance of a coherence length. The NS interface 
connects electrons and holes of the same channel with a reflection amplitude 
depending on the reduced chemical potential
$\mu_{\nu}=\mu-\hbar^2\bv{k}_{\bot}^2/2m$. In the limit
$\epsilon,\Delta \ll \mu_{\nu}$, the BdG equations are simplified by
linearizing the dispersion relation around the effective Fermi wave number
$k_{\nu}^{\left(0\right)}=\sqrt{2m\mu_{\nu}/\hbar}$. The boundary conditions 
are fulfilled in the Andreev approximation, which treats the wave number 
$k_{\nu}$ in the phase factor of the single-particle excitations only to 
zeroth order. As a consequence, incoming electrons are purely reflected into
holes and vice versa. The reflection of electron (hole) channels into
hole (electron) channels is thus described by scalar reflection
amplitudes, which turn out to depend only on the energy $\epsilon$ and
not on the channel index. The reflection matrix at the NS interface is
given by
\begin{equation} \hat{r}_{I}\left(\epsilon\right)= \left(\begin{array}
{cc} 0 & r_{he}\left(\epsilon\right) \\ r_{eh}\left(\epsilon\right) & 0 
\end{array}\right) = \left(\begin{array}{cc} 0 & e^{-i\chi}\Gamma
\left(\epsilon\right)\\ e^{i\chi}\Gamma\left(\epsilon\right) & 0 
\end{array}\right), \end{equation}
with $\Gamma\left(\epsilon\right)$ defined by
\begin{equation} \Gamma\left(\epsilon\right)=\left\{\begin{array}{ll}
\frac{{\textstyle\epsilon-\mbox{sign}\left(\epsilon\right)
\sqrt{\epsilon^2-\Delta^2}}}{{\textstyle \Delta}} \,\sim \frac{{\textstyle 
\Delta}}{{\textstyle 2|\epsilon|}}, & |\epsilon| > \Delta, \\ 
\frac{{\textstyle \epsilon-i\sqrt{\Delta^2-\epsilon^2}}}{{\textstyle \Delta}}
 = \exp\left({\textstyle -i\arccos\frac{{\textstyle \epsilon}}{\Delta}}\right), 
& |\epsilon| < \Delta. 
\end{array} \right. \label{Gamma} \end{equation}
The $2N\times 2N$ global reflection matrices $R_{ee}$ and $R_{he}$ 
can be determined from equation (\ref{Rhat}), and using (\ref{cond}), we 
obtain the multichannel spectral conductance formula
\begin{eqnarray} G_s\left(\epsilon,V\right)&=&\frac{2e^2}{h}\left(1+\mid
\Gamma\left(\epsilon\right)\mid^2\right)   \mbox{Tr}\left\{t_{21}^{\dag}
\left(\epsilon\right)\left[1-\Gamma^*\left(\epsilon\right)^{2}r_{22}^{
\top}\left(-\epsilon\right)r_{22}^{\dag}\left(\epsilon\right)\right]^{-1}
 \right. \nonumber \\ &&\left. \times\Big(1-\mid\Gamma\left(\epsilon
\right)\mid^2 r_{22}^{\top}\left(-\epsilon\right)r_{22}^*\left(-\epsilon
\right)\Big) \left[1-\Gamma\left(\epsilon\right)^2 r_{22}\left(\epsilon
\right)r_{22}^*\left(-\epsilon\right)\right]^{-1}  t_{21}\left(\epsilon
\right)\right\}, \label{speccond} \end{eqnarray}
valid at all energies. Eq. (\ref{speccond}) is one of the central results of 
this work. Combined with equation (\ref{current}), it provides the finite 
voltage, finite temperature CVC of a disordered normal--superconducting
junction in the Andreev approximation. The spectral conductance
depends on the scattering matrices of the electrons at energies $\pm
\epsilon$ as a signature of the presence of Andreev reflection (the
explicit voltage dependence, although present in both reflection and
transmission matrices, is not indicated in the notation). The dependence 
of this formula on the phases of the reflection and transmission amplitudes
proves crucial in determining the resonance peaks in the conductance. 
The elementary process, which contributes to these phases is the propagation 
of an electron and a hole between the disorder region and the NS interface.

If no inter-channel mixing takes place, i.~e., the matrices $t_{ij}$ and
$r_{ii}$ are diagonal, the conductance reduces to the quasi one-dimensional 
form,
\begin{equation} G_s\left(\epsilon,V\right)=\frac{2e^2}{h}\sum_{n=1}^{N}
 \frac{\left(1+\mid\Gamma\left(\epsilon\right)\mid^2\right)\quad T_{\nu}
\left(\epsilon,V\right)\left[1-\mid\Gamma\left(\epsilon\right)\mid^2 
R_{\nu}\left(-\epsilon,V\right)\right]}{1+\mid\Gamma\left(\epsilon\right)
\mid^4 R_{\nu}\left(\epsilon,V\right) R_{\nu}\left(-\epsilon,V\right)- 2 Re
\left[\Gamma\left(\epsilon\right)^2 r_{\nu}\left(\epsilon,V\right) r_{\nu}^*
\left(-\epsilon,V\right)\right]}. \label{onechannel} \end{equation}
$r_{\nu}=\left(r_{22}\right)_{\nu\nu}$ is the reflection amplitude for a
state coming in from the right side of the scattering region (Fig. 1) and
$R_{\nu}=|r_{\nu}|^2$ and $T_{\nu}=1-R_{\nu}$ denote the reflection and 
transmission probabilities of the $\nu$-th channel. The last
term in the denominator describes the crucial scattering process which 
involves twice the propagation between the disorder region and the NS 
interface, once as an electron and a second time as a hole. 

For voltages well {\it above} the gap, $|\epsilon|, V\, \gg\Delta$ (still 
assuming $|\epsilon|\ll \mu$), the Andreev reflection is strongly 
suppressed and drops according to 
$\Gamma\left(\epsilon,V\right)\sim \Delta/2|\epsilon|\,\to 0$. The spectral 
conductance (\ref{speccond}) asymptotically approaches the expression for
a normal junction,
\begin{equation} G_s\left(\epsilon,V\right)=\frac{2e^2}{h}\mbox{Tr}
\left\{t_{21}^{\dag}\left(\epsilon,V\right)t_{21}\left(\epsilon,V\right)
\right\} \label{land} \end{equation}
(note that this Landauer type formula is the spectral conductance at finite 
voltage and energy and deviates from the differential conductance). The 
conductance (\ref{land}) exhibits no particular symmetry properties with 
respect to sign reversal of the bias. 

At voltages {\it below} the gap, we make use of $\mid\Gamma\left(\epsilon
\right)\mid=1$ and $T_{\nu}=1-R_{\nu}$ and obtain the spectral conductance 
(\ref{onechannel}) in the form,
\begin{equation} G_s\left(\epsilon,V\right)=\frac{4e^2}{h}\sum_{n=1}^{N} 
\frac{ T_{\nu}\left(\epsilon,V\right) T_{\nu}\left(-\epsilon,V\right)} 
{1+ R_{\nu}\left(\epsilon,V\right) R_{\nu}\left(-\epsilon,V\right)- 
2 Re\left[\Gamma \left(\epsilon\right)^2 r_{\nu}\left(\epsilon,V\right) 
r_{\nu}^*\left(-\epsilon,V\right)\right]}. \label{subonechannel} 
\end{equation}
The reflection and transmission coefficients at $\pm\epsilon$ are 
symmetrically involved in this formula, which results in the symmetry of 
the CVC discussed in section II. In contrast, the spectral conductance 
(\ref{onechannel}) at voltages above the gap becomes increasingly asymmetric 
as it asymptotically approaches the Landauer expression. As an important
difference to the normal expression (\ref{land}), the conductance 
(\ref{subonechannel}) contains the phase information of the scattering
processes, which is imperative for the distinction of zero and finite bias
peaks in the double barrier NS junction dicussed below.

The linear response limit ($\epsilon,V\,\to 0$) of (\ref{subonechannel}) 
can be determined using $\Gamma\left(0\right)^2=1$ and $R_{\nu}=1-T_{\nu}$ 
and takes the form\cite{beenakker}
\begin{equation} G\left(0\right)=\frac{4e^2}{h}\sum_{n}\frac{T_{\nu}\left(0
\right)^2}{\left(2-T_{\nu}\left(0\right)\right)^2}. \label{been} 
\end{equation}
Eq. (\ref{been}) has proved very useful through its remarkable simplicity.

\section{Double barrier NS junction}

In this section we apply the above results to a double barrier NINIS junction, 
which is a model system for the study of the interplay between normal- and 
Andreev- levels in a Fabry--Perot type I$_1$NI$_2$ interferometer. 
We will learn about the mechanism govering the appearance of zero and finite 
bias anomalies and give insight into the interpretation of the corresponding 
anomalies in a dirty NS junction. First, we discuss the structure in the 
conductance of a single channel NI$_1$NI$_2$S junction, which we trace back to 
the presence of Andreev resonances.
Secondly, we present numerical results for a multichannel junction, showing
that the typical resonance structure of a single channel survives the 
summation over the channels. This stability is a peculiarity of the 
superconducting system not observed in a normal NI$_1$NI$_2$N double barrier 
junction. 

Since the channels separate in the double barrier problem, we can make use 
of the result (\ref{onechannel}) for the conductance $G_s$\cite{index}. 
$G_s$ depends on 
the phases $\varphi\left(\pm\epsilon\right)$ of the reflection amplitudes 
$r\left(\pm\epsilon\right)$ as well as on the complex amplitude 
$\Gamma\left(\epsilon\right)$ of the Andreev reflection. We use the notation
$T\left(\pm\epsilon\right)=T_{\pm}$, $R\left(\pm\epsilon\right)=R_{\pm}$, and
$r\left(\pm\epsilon\right)=\sqrt{R_{\pm}}e^{i\varphi\left(\pm\epsilon
\right)}$ for the reflection amplitude, the phase factors being determined 
by the potential barriers $I_1$ and $I_2$ and the propagation between them 
(for simplicity, we omit the explicit dependence of the scattering matrix on 
voltage). We rewrite the amplitude of the Andreev reflection as
$\Gamma\left(\epsilon\right)=|\Gamma| e^{-i\vartheta\left(\epsilon\right)}$
with the phase $\vartheta\left(\epsilon\right)= \arccos\left(\epsilon/\Delta
\right)$ below the gap and vanishing above. The conductance simplifies to
\begin{equation} G_s\left(\epsilon\right)=\frac{2\left(1+|\Gamma|^2\right)e^2}
{h} \frac{ T_+ \,\, \left(1-|\Gamma|^2 R_-\right)}{1 + |\Gamma|^4 R_+ R_-  
-2|\Gamma|^2\sqrt{R_+R_-} \cos\left[\varphi\left(\epsilon\right) -\varphi
\left(-\epsilon\right) -2\vartheta\left(\epsilon\right)\right] }, 
\label{advert} \end{equation}
which is always less or equal to the universal value $4e^2/h$. Note that 
the Andreev reflection is supressed above the gap ($|\Gamma| < 1$), while
the phase $\vartheta\left(\epsilon\right)$ vanishes.\\
Below the gap, the phase $\vartheta\left(\epsilon\right)$ is decisive for the 
resonances produced by the interference of multiple scattering processes.
The maximal conductance (which is twice the normal conductance $2e^2/h$) is 
only assumed if the reflection probabilities $R_{\pm}$ are equal and the 
phase $\varphi\left(\epsilon\right)$ fulfills the resonance condition,
\begin{equation} \cos\left[\varphi\left(\epsilon\right) -\varphi\left(
-\epsilon\right) -2\vartheta\left(\epsilon\right)\right]=1. \label{res} 
\end{equation}
The analogous property is found in a double barrier system within a normal
constriction (NI$_1$N$_2$N), where the maximal conductance $2e^2/h$ is reached
if the reflection probabilities of the two barriers coincide at the resonance
energy.
 
Using the conductance formula (\ref{advert}), we consider a one channel NINS 
junction, which consists of a ballistic NS junction containing a single 
potential barrier at a distance $d$ from the perfect NS interface. 
In the limit of a high potential barrier, $R_+$ and $R_-$ are approximately 
equal. The reflection amplitudes $r\left(\pm\epsilon\right)$ describing the 
propagation of electrons and holes
have nearly constant modulus and phases $\varphi\left(\pm\epsilon\right)=
\pi + 2k_{\pm}d$. Using the wave number $k_{\pm}=mv_F\pm\epsilon/v_F$ 
($v_F$ is the Fermi velocity of the channel), 
the resonance condition (\ref{res}) yields the spectrum of Andreev levels,
\begin{equation} \epsilon_{n}=\frac{v_F}{2d}\left(n\pi+\arccos 
\frac{\epsilon_{n}}{\Delta} \right), \label{alevel} 
\end{equation}
which predicts resonances in the conductance of a typical width proportional
to the transmission $T$ of the barrier (similar Andreev resonances are found 
for voltages above the gap at $\epsilon_{n}=n\pi v_F/2d$ with a width 
roughly $\propto T/\Gamma\left(\epsilon\right)$). The phase $\vartheta\left(
\epsilon\right)$ varies between $-\pi/2$ and $0$ from $\epsilon=0$ to 
$\epsilon=\Delta$ and guarantees the existence of at least one 
Andreev resonance for arbitrarily small $d$. In the $d\to 0$ limit, 
the weight of this resonance lines up with the gap voltage and we recover 
the NIS junction as discussed by Blonder et al.\cite{btk}. featuring a 
suppressed subgap conductivity and a peak in the differential conductance at 
the gap voltage. This peak can be understood in terms of the Andreev 
resonance which moves to the gap energy for $d\to 0$.

We now introduce an additional barrier at the NS interface and analyze the 
resulting double barrier NI$_1$NI$_2$S junction using the same 
conductance formula (\ref{advert}). Following our definitions, 
$\varphi\left(\pm\epsilon\right)$ represent the phases for the reflection  
of electrons entering the double barrier scattering region from the right
(superconducting side). The corresponding reflection amplitudes are given by
\begin{equation} r\left(\pm\epsilon\right)=r_2 + \frac{t_2^2 r_1 e^{2ik_{\pm}d}}
{1-r_1 r_2 e^{2ik_{\pm}d}}, \label{rpm} \end{equation}
where $r_i, t_i$ are the amplitudes of the left ($i=1$) and right ($i=2$) 
barrier. The phase of this reflection amplitude plays the major role in 
determining the structure of the conductance, as it controls the existence 
of resonances according to (\ref{res}). Let us fix the barrier I$_1$ and 
increase 
I$_2$ slowly, keeping their strengths I$_1>$ I$_2$. In this situation, the INI 
interferometer develops pronounced Andreev resonances. For $r_1\gg r_2$, the 
phase $\varphi\left(\pm\epsilon\right)$ of the reflection amplitude
$r\left(\pm\epsilon\right) \approx t_2^2r_1e^{2ik_{\pm}d}$ produces a linear 
energy dependence of the phase $\varphi\left(\epsilon\right)$, which changes by
$2\pi$ on the scale $v_F/d$ and results in equidistant resonances\cite{note},
 in accordance with (\ref{alevel}). As the strength of I$_2$ 
is increased, the resonances pair up as is 
illustrated in Figure 2. The phase function $\varphi\left(\epsilon\right)$, as 
displayed in Figure 2 (solid line), can be used to determine the location of 
the resonances by finding those combinations of energies $\pm\epsilon$, which 
have a phase difference\cite{note} $\Delta\varphi\left(\epsilon\right)=
\varphi\left(\epsilon\right)-\varphi\left(-\epsilon\right)=\pi+2n\pi$. 
The double period of $\Delta\varphi\left(\epsilon\right)$ with respect to the 
period of $\varphi\left(\epsilon\right)$ accounts for the pairing of the 
resonances. 

When the strengths of the barriers become of the same order, I$_1\sim$ I$_2$, 
the spectral weight of the INI interferometer is shared by Andreev quasi-bound 
states of a mixed electron--hole character and  normal electron quasi-bound 
states. Due to the large gradient of the phase close 
to the normal resonances, see Fig.~2, the Andreev resonances tend to be pinned 
to normal resonances at either $+\epsilon$ or $-\epsilon$ (a notable exception 
to this rule is found when the normal resonance is aligned with the Fermi 
energy, 
but the Andreev resonances remain at finite bias). While the Andreev bound 
states contribute to the current, normal bound states do not couple to the 
superconductor and thus do not participate in the charge transport. This is 
reflected by the symmetry of the $G_s\left(\epsilon\right)$ under reversal of 
voltage which is is observed for all barrier strengths, see Figure 2.

As the barrier strength is increased further to I$_2>$ I$_1$, the Andreev 
resonances are weakened and eventually disappear. Although the normal 
resonances dominate in the INI interferometer in this regime, only the weak
Andreev resonances show up in the conductance, which thus exhibits only a weak 
(but still symmetric) subgap structure. The phase function 
$\varphi\left(\epsilon\right)$ in 
Figure 2 (dashed line) becomes nearly constant for $r_1 \ll r_2$, 
 see (\ref{rpm}), and the phase condition for resonance (\ref{res}) cannot be 
met.

Let us compare the transport in the double barrier systems NI$_1$NI$_2$S and 
NI$_2$NI$_1$S, i.e., with inverse sequences of the barriers I$_1$ and I$_2$. 
Note that the transmission $T\left(\epsilon\right)$ is identical for both 
cases, 
see Figure 2 (dotted line). Both the (finite voltage) conductance in the 
normal NININ junction (\ref{land}) as well as the linear response conductance 
in the superconducting NINIS junction (\ref{been}) are thus independent of the 
sequence of the barriers I$_1$ and I$_2$. Let us assume that I$_1 \gg$ I$_2$. 
In this first barrier sequence, we have a strong energy dependence of the phase 
$\varphi\left(\epsilon\right)$ (solid line in Fig.\ 2), 
which implies the existence of Andreev type resonances at 
finite bias. The electrons entering the INI interferometer from the normal lead,
are given enough time to build up an Andreev resonance and preferably leave into
the superconductor. For the inverse barrier sequence, the barrier I$_2$ at the
NS interface dominates. The weak energy dependence of the phase 
$\varphi\left(\epsilon\right)$ of reflection allows no sharp 
resonances to build up (dashed line in Fig.\ 2). This reflects the fact that 
the electrons which enter the INI region leave through I$_1$ into the normal 
lead before an Andreev resonance can build up.
In summary, the spectral density in the INI interferometer changes radically
with the coupling strengths of the normal and superconducting leads. Normal
resonances dominate when the interferometer is coupled more strongly to the
normal lead, whereas Andreev resonances take over in weight when the coupling
to the superconductor is stronger. At any instant, however, only the Andreev 
states participate in the charge transport.

We turn to the numerical analysis of a multichannel NINIS junction which we
carry out using the conductance formula (\ref{onechannel}). This formula allows 
to extend the linear response study of Ref. \cite{melsen} to finite voltage and 
temperature. We investigate an NI$_1$NI$_2$S junction with two 
$\delta$-function 
barriers of typical strength $H=\int V\left(x\right) dx \approx \hbar v_F$ and 
corresponding reflection probability $R=H^2/\left(H^2+\hbar^2v_F^2\right)$ 
assuming values between $R=0.2$ and $R=1$. We vary the relative barrier 
strengths 
to cover the range between the two limits I$_1>$ I$_2$ and I$_1<$ I$_2$ 
discussed above. 
The distance between the barriers is chosen to be of the order of or larger than
the coherence length of the superconductor, providing the forward channel with
one to a few Andreev resonances. The number of resonances increases 
with the incidence angle of the channels. We chose leads with a cross section 
area of $\left(100/k_F\right)^2$, which amounts to about $800$ transverse 
channels. The ratio of the energy gap to the Fermi energy is assumed to be 
$\Delta/\epsilon_F=0.002$. Each channel features the typical conductance
structure of paired Andreev resonances exposed above. Their positions and 
widths 
depend on the ratio of the barrier strengths I$_1$ and I$_2$ as well as the 
longitudinal kinetic energy of the single channels. Remarkably, the overall 
conductance,
which is obtained through summation of single channel conductances, exhibits a 
characteristic subgap structure signalling the presence of Andreev resonances.
In contrast, the overall conductance of the corresponding NI$_1$NI$_2$N 
junction 
is practically constant, the normal resonances of the INI region having been 
averaged out.

The numerical study of a 3-dimensional NINS junction has shown that both the 
positions and the number of resonances in the overall conductance correspond to
those in the forward channel\cite{bagwell,angle}. In the NINIS junction, we do 
not find a direct correspondence of the resonances of the total conductance 
with 
the forward channel nor with any other specific channel, although a clear 
resonance structure still survives the summation over the channels.

Let us concentrate on the conductance formula (\ref{subonechannel}), valid 
at subgap voltages, and on the properties of the CVC close to zero voltage. 
For I$_1>$ I$_2$, the denominator of (\ref{subonechannel}) changes rapidly with 
the strong energy dependence of the phase of $r_{\nu}\left(\epsilon\right)$ 
which is responsible for the appearance of conductance peaks at
finite voltage. The pronounced structure in the conductance survives the 
summation over the channels as displayed in Figures 3 and 4 (solid lines). 
The repulsion of the Andreev levels
around zero voltage produces a minimum in $dI/dV$ at zero voltage. For 
I$_1<$ I$_2$, the phase of the reflection amplitudes $r_{\nu}\left(\epsilon
\right)$ has negligible energy dependence and the numerator of 
(\ref{subonechannel}) dictates the features of the conductance. The expansion 
of the product 
$T_{\nu}\left(\epsilon\right)T_{\nu}\left(-\epsilon\right)=T_{\nu}^2- 
\epsilon^2 T_{\nu}'^2$ about zero energy shows the existence of a zero bias 
maximum (the denominator can be seen not to alter this property as long as the 
total transmission of the double barrier system is not too large, 
$T_{\nu}<0.55$). 
The zero bias anomaly shows up as a characteristic property of the overall 
conductance, see Figures 3 and 4 (dotted lines). The zero bias maximum 
coincides 
with the maximum of the conductance product $G\left(\epsilon\right)G\left(-
\epsilon\right)$ of the corresponding NININ junction at zero energy. Figures 3 
and 4 illustrate the crossover from zero to finite bias anomalies for two 
different 
interbarrier distances $d$ as the strength of barrier I$_1$ is increased and
I$_2$ is kept fixed. For an interbarrier distance larger than the coherence 
length of the superconductor, several Andreev resonances show up (see Fig.~4). 
Note that the inversion of the barrier sequence transforms the zero
voltage conductance from a local minimum to a local maximum, while keeping the
same zero voltage conductance. This is illustrated in Fig.~4 by the pair of 
solid lines around zero voltage corresponding to inverse barrier sequences.

In an attempt to understand the width of zero and finite bias peaks as well as 
the position of the first finite bias anomaly, we have compared them to the 
Thouless energy of the system. The Thouless energy\cite{gefen} $E_c$ in a 
disordered system can be defined as the product of the dimensionless 
normal state conductance $g$ and the level spacing $\delta E$, 
$E_c=g \delta E$. 
We have determined a corresponding Thouless energy in our system by
taking the level spacing of the INI box multiplied by its overall conductance. 
In the numerical simulations of a weakly transparent barrier system, 
we found good agreement of this energy scale with both the width and the 
position 
of the finite bias anomaly. The width of the zero bias anomaly 
coincided with the characteristic energy scale of the conductance correlator 
$\langle G\left(E+\epsilon\right)G\left(E\right)\rangle_E$ as well as with 
$E_c$. 
In this case, the distribution of the channel transmission of 
the double barrier system is bimodal and resembles strongly to that of a 
dirty system\cite{melsen}. As the overall transparency
approaches $1$, however, the width of the resonances and $E_c$ disagree. 
In this limit, the double barrier system does not 
represent the bimodal distribution of the dirty system properly and we cannot 
expect the above definition of the Thouless energy to be a sensible quantity. 
Of experimental interest is the effect of temperature on the conductance
resonances predicted above. A finite bias anomaly will be smeared out to 
a zero bias anomaly for a temperature of the order of the Thouless energy. In
the systems investigated here this energy is of the order of $0.1\Delta$.

The interference of multiple scattering processes between the scattering 
region and the NS interface thus produces an interesting structure in the 
differential conductance. Recently, experiments in disordered
NS and SmS junctions\cite{kastalsky,nguyen,nitta,bakker,magnee} have
concentrated on the observation of zero and finite bias anomalies. It has been
understood theoretically\cite{volkov,marmokos,yip,wees} that these features are
due to the interplay between the barrier at the NS interface and
disorder in the normal lead. At small disorder, the differential conductance 
exhibits a zero bias maximum, while at large enough disorder, a finite bias 
peak 
is expected\cite{yip} at a voltage of the order of the Thouless energy $E_c$ of 
the normal lead. This has recently been confirmed in an 
experiment\cite{sanquer}. 
Here, we have found the existence of the analog features in a ballistic double
barrier NS junction (in  addition, the double barrier junction shows higher 
harmonics in the resonances of the conductance). The ballistic point of view 
applied to the disordered NS junction thus suggests the interpretation of the 
finite bias anomaly as a superposition of resonances due to 
quasi-bound Andreev states between the superconductor and the disorder.

\section{Conclusion}

We have used the scattering matrix approach including finite voltage and 
nonzero temperature dependences to describe the current--voltage 
characteristics 
of NS junctions. The current--voltage relation has been expressed through a 
spectral conductance which takes into account the explicit voltage dependence 
of the scattering problem. 
We have demonstrated the existence of a symmetry of the spectral conductance 
under sign reversal in the subgap regime. Above the gap, the symmetry is
destroyed as the conductance approaches the Landauer expression. We have 
presented the spectral conductance formula in the Andreev approximation 
for a multichannel NS junction at all voltages in (\ref{speccond}). This 
result has enabled us to carry out a study of a double barrier NINIS junction 
at finite voltage, which has revealed the crossover from normal to Andreev 
resonances in the INI interferometer as the ratio of the barrier strength is
varied. Interestingly, a radical dependence of the conductance on the sequence 
of
the barriers has been found. We have learned about the mechanism governing the 
occurrence of and the crossover between zero and finite bias anomalies, tracing 
them back to the energy dependence of the modulus and the phase of the 
reflection amplitude, respectively. Finally, we have established the similarity
of the conductance of the double barrier NS junction with a disordered NS 
junction.

We are indebted to M.\ Sanquer and W.\ Poirier for numerous
stimulating discussions. We wish to acknowledge an enlightening
discussion with Y.\ Imry. A.~L.~F.\ is grateful to A.\ van Otterlo for
his competent support. The work of G.~B.~L.\ was done during his stay
at SPEC, Saclay, and partly at ETH-Z\"urich, the hospitality of which
is greatly appreciated. We acknowledge the financial support of the
Schweizerische Nationalfonds zur F\"orderung der wissenschaftlichen 
Forschung. G.~B.~L.\ acknowledges partial support by the NATO collaboration
research program through the grant N. 921333.

\begin{figure}[h]
\centerline{\psfig{figure=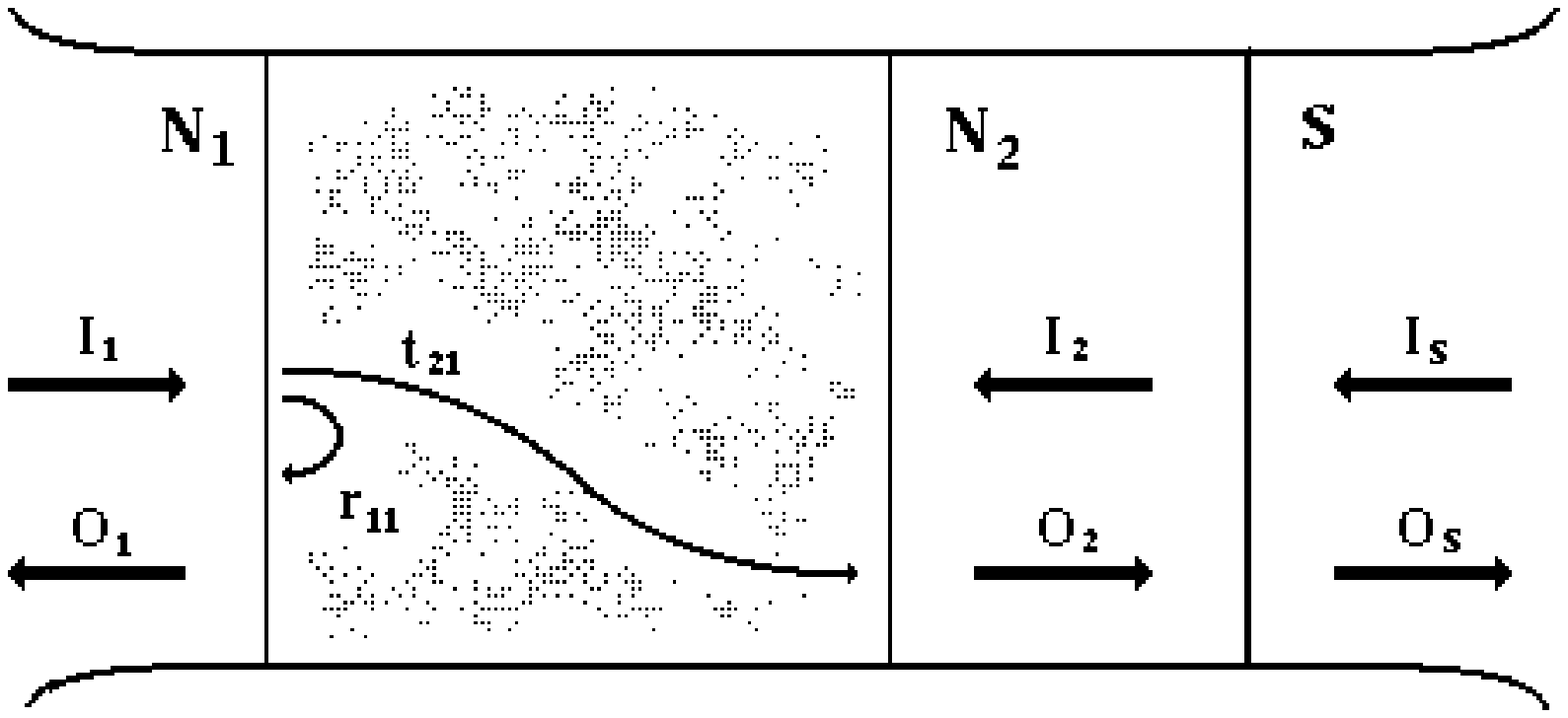,height=5cm}}
\caption{Schematic structure of a disordered NS junction. 
  Ballistic normal ($N_1$, $N_2$) and superconducting leads (S) are
  coupled to reservoirs at chemical potential $\mu-eV$ and $\mu$,
  respectively. Scattering is limited to the hatched region between
  the ballistic leads $N_1$ and $N_2$.}
\end{figure}

\pagebreak

\begin{figure}[h]
\noindent
\centerline{\psfig{figure=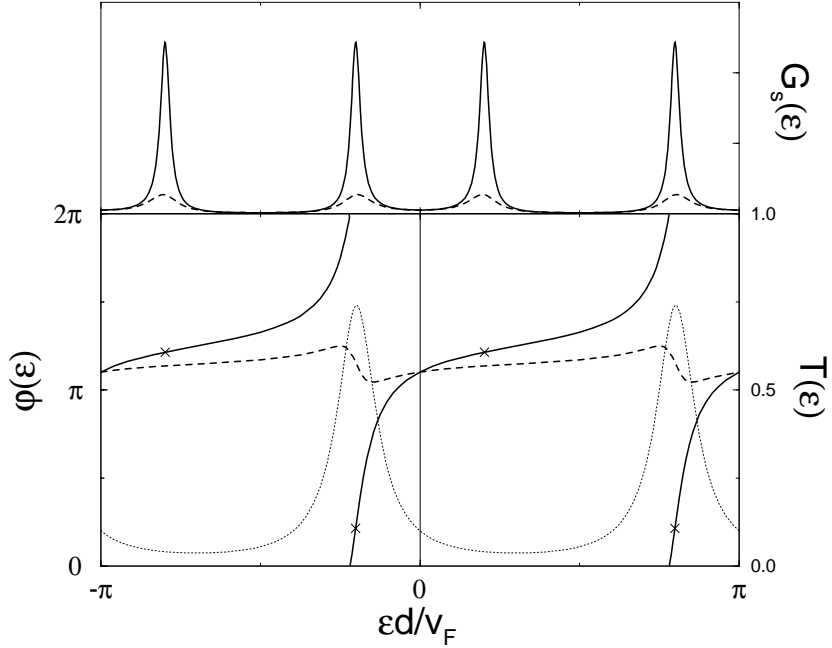,height=9cm,angle=-90}}
\vspace{1cm}
\caption{Andreev resonances and resonance condition for the phase $\varphi$. 
Bottom: phase $\varphi\left(\epsilon\right)$ of the reflection amplitude versus
energy. The solid line represents the I$_1$NI$_2$ interferometer with barriers 
strengths $H_1=2\hbar v_F$ ($R_1=0.8$) and $H_2=\hbar v_F$ ($R_2=0.5$), the 
dashed line stands for the inverse barrier sequence ($R_1=0.5$ and $R_2=0.8$). 
The {\it Andreev} resonance condition for the phase is met for a pair of 
energies $\pm\epsilon_n$ with phase difference 
$\Delta\varphi\left(\epsilon_n\right)=\pi+2n\pi$.
This phase condition can be fulfilled only by the first barrier sequence 
($R_1>R_2$, solid line) at those energies indicated in the graph, and 
produces a peak in the conductance of the NINIS structure. 
Interchanging the strengths of the two barriers,
the {\it Andreev} resonances are greatly reduced, although the NININ double 
barrier system has the same overall transmission probability $T$ (dotted line) 
as determined by the {\it normal} resonances.
Top: conductance (arbitrary units) of the double barrier NS junction 
versus energy, the solid line again representing the barriers 
$R_1=0.8$, $R_2=0.5$, and the dashed line the barriers $R_1=0.5$, $R_2=0.8$.
Note the symmetry of the resonances with respect to $\epsilon=0$, which is
due to their electron-hole character.}
\end{figure}

\begin{figure}[h]
\noindent
\centerline{\psfig{figure=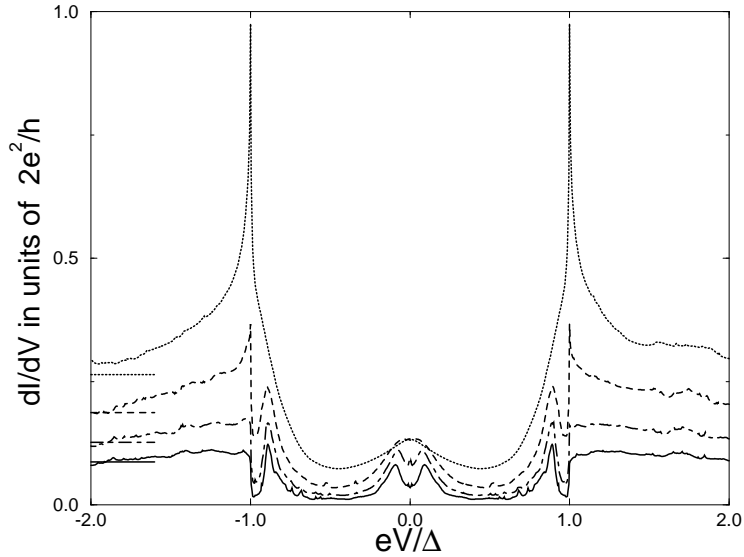,angle=-90,height=9cm}}
\vspace{1cm}
\caption{Differential conductance for a multichannel NINIS junction with 
interbarrier distance $d=2v_F/\Delta=2\pi\xi$. The average conductance per 
channel is plotted 
versus voltage at temperature $T=0$. The barrier strengths $H$ (reflection 
probabilities $R=(1+H^2)/H^2$) in units of $\hbar v_F$ are $H_1=0.5$ 
($R_1=0.2$) (dotted line), $1.0$ ($0.5$) (dashed), $1.5$ ($0.7$) (dot-dashed), 
and $2.0$ ($0.8$) (solid) in the normal region, while $H_2=1$ ($R_2=0.5$) 
at the NS interface. The normal state conductance, which is roughly 
independent over this voltage range, is indicated on the left. With increasing
barrier strength $H_1$ the zero bias anomaly develops into a finite bias anomaly
as the Andreev resonance is formed for $H_1>H_2$.}
\end{figure}

\pagebreak

\begin{figure}[h]
\noindent
\centerline{\psfig{figure=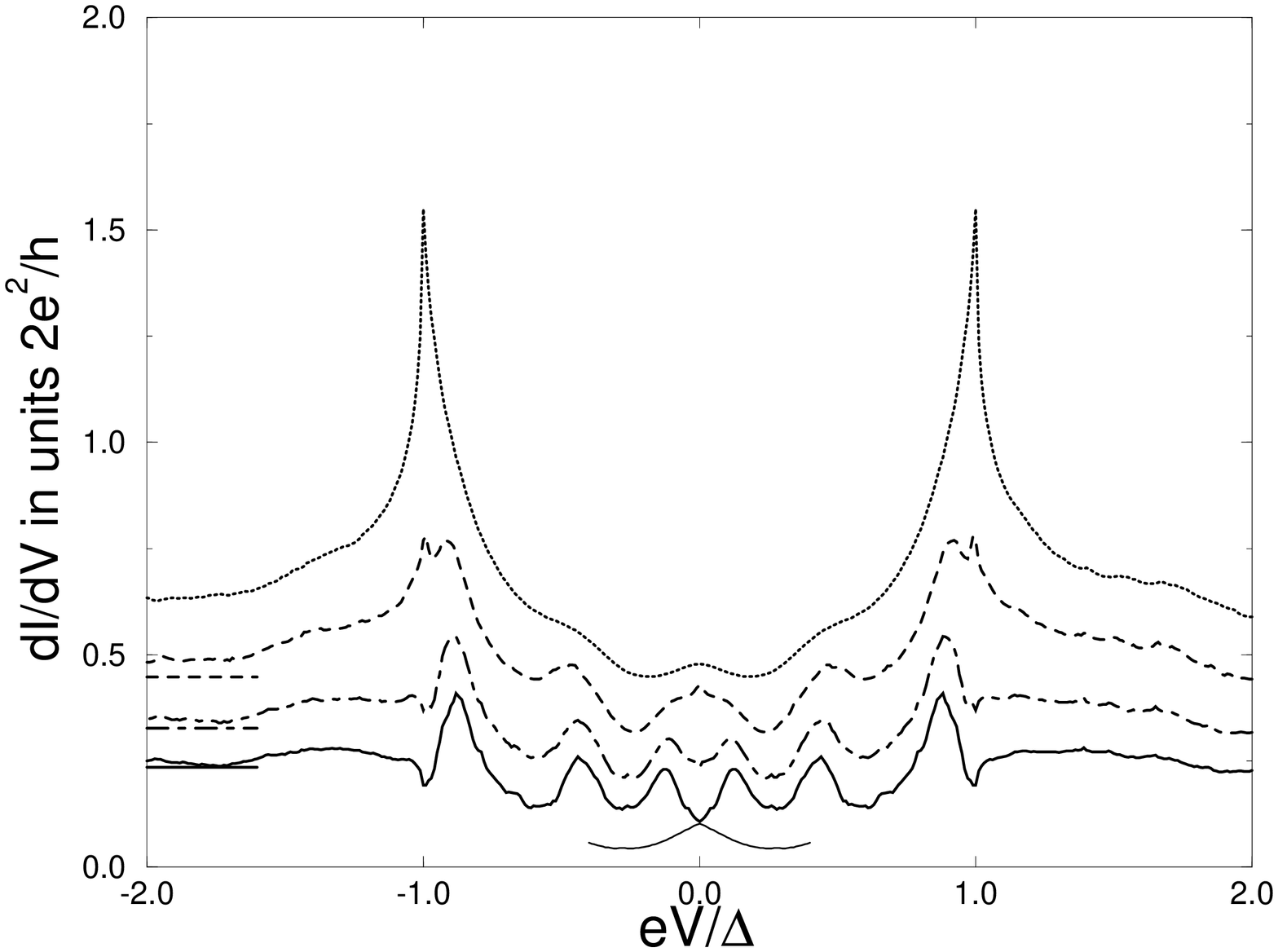,angle=-90,height=9cm}}
\vspace{1cm}
\caption{Differential conductance for a multichannel NINIS junction of 
width $d=4v_F/\Delta=4\pi\xi$. The average conductance per channel is plotted 
versus voltage for temperature $T=0$. The barrier strengths (reflection
probabilities) in units of $\hbar v_F$ are $H_1=0.2$ ($R_1=0.04$) 
(dotted line), $0.5$ ($0.2$) (dashed), $0.8$ ($0.4$) (dot-dashed), 
and $1.1$ ($0.54$) (solid) in the normal region, and fixed at $H_2=0.5$ 
($R_2=0.2$)
at the NS interface. The normal state conductivity is indicated on the left.
As the strength of $H_1$ is increased, the zero bias anomaly
turn into an finite bias anomaly and several Andreev resonances appear.
For the last choice of barriers $H_1=1.1$ and $H_2=0.5$ we have interchanged
the barrier sequence: the conductance at zero voltage remains the same while 
changing from a zero bias minimum to a zero bias maximum (short solid line).}
\end{figure}


\begin{thebibliography}{99}

\bibitem{deGennes}P.~G.~ de Gennes, Superconductivity of Metals and
  Alloys (Benjamin, New York, 1966).  \bibitem{andreev}A.~F.~Andreev,
  Sov.\ Phys.\ JETP {\bf 19}, 1228 (1964).
\bibitem{btk}G.~E.~Blonder, M.~Tinkham, and T.~M~Klapwijk, Phys.\ 
  Rev.\ B {\bf 25}, 4515 (1982).  
\bibitem{beenakker}C.~W.~Beenakker, Phys.\ Rev.\ B {\bf 46}, 12841
  (1992).
\bibitem{landauer}R.\ Landauer, Philos.\ Mag.\ {\bf 21}, 863 (1970).
\bibitem{buttland}M.~B\"uttiker, Y.~Imry, R.~Landauer, and S.~Pinhas,
  Phys.\ Rev.\ B {\bf 31}, 6207 (1985).
\bibitem{pothier}H.~Pothier {\em et
    al.}, Phys.\ Rev.\ Lett.\ {\bf 73}, 2488 (1994).
\bibitem{petrashov}V.~T.~Petrashov {\em et al.}, Phys.\ Rev.\ Lett.\ 
  {\bf 74}, 5268 (1995).  \bibitem{pannetier}H.~Courtois {\em et al.},
  Phys.\ Rev.\ Lett.\ {\bf 76}, 130 (1996).
\bibitem{volkov}A.~F.~Volkov, A.~V.~Zaitsev, and T.~M.\ Klapwijk,
  Physica C {\bf 210}, 21 (1993).  \bibitem{marmokos}I.~K.~Marmokos,
  C.~W.~Beenakker, and R.~A.~Jalabert, Phys.\ Rev.\ B, {\bf 48}, 2811
  (1993).  \bibitem{yip}S.~Yip, Phys.\ Rev.\ B, {\bf 52}, 15504
  (1995).  \bibitem{kastalsky}A.~Kastalsky {\em et al.}, Phys.\ Rev.\ 
  Lett.\ {\bf 67}, 3026 (1991).  \bibitem{nguyen}C.~Nguyen {\em et
    al.}, Phys.\ Rev.\ Lett.\ {\bf 69}, 2847 (1992).
\bibitem{nitta}J.~Nitta {\em et al.}, Phys.\ Rev.\ B {\bf 49}, 3659
  (1994).  \bibitem{bakker}S.~J.~M.~Bakker {\em et al.}, Phys.\ Rev.\ 
  B {\bf 49}, 13275 (1994).  \bibitem{magnee}P.~H.~C.~Magn\'ee {\em et
    al.}, Phys.\ Rev.\ B {\bf 50}, 4594 (1994).
\bibitem{sanquer}W.~Poirier, D.~Mailly, and M.~Sanquer, preprint.
\bibitem{nazarov}T.~H.~Stoof and Yu.~V.~Nazarov, Phys.\ Rev.\ Lett.\ 
  {\bf 76}, 823 (1996); T.~H.~Stoof and Yu.~V.~Nazarov, Phys.\ Rev.\ B
  {\bf 53}, 14496 (1996).  \bibitem{hekking}F.~W.~J.~Hekking and
  Yu.~V.~Nazarov, Phys.\ Rev.\ B {\bf 49}, 6847 (1994).
\bibitem{kulik}I.~O.~Kulik, Sov.\  Phys.\ JETP {\bf 30}, 944 (1970).
\bibitem{lambert}C.~J.~Lambert, J.\ Phys.\ Cond.\ Mat.\ {\bf 3}, 6579
  (1991); C.~J.\ Lambert, V.~H.\ Hui, and S.~J.\ Robinson, 
  J.\ Phys.\ Cond.\ Mat.\ {\bf5}, 4187 (1993). 
\bibitem{takane}Y.~Takane and H.~Ebisawa, J.\ Phys.\ Soc.\ 
  Jap. {\bf 61}, 1685 (1991); {\bf 61}, 2858 (1992).
\bibitem{brouwer}P.~W.~Brouwer, C.~W.~J.~Beenakker, Phys.\ 
  Rev.\ B {\bf 52}, R3868 (1995).  
\bibitem{elhole}Electron-hole
  symmetry is however not required for any of the properties derived
  below, and will not be assumed further.  \bibitem{prob}For the
  definition of probability and charge current see Ref. \cite{btk}.
\bibitem{probcharge}Note that the connection the unit charge currents on
either side of the interface will not produce unitary, since quasiparticle 
charge current is not a conserved quantity in the superconductor.
\bibitem{net}This can be shown using the unitarity of the global scattering
matrix including the Andreev process by an extension of the argument given for 
a normal junction.
\bibitem{lamblead}M.~Leadbeater and C.~J.~Lambert, J.\ Phys.\ Cond.\ Mat.\
  {\bf 8}, L345 (1996). 
\bibitem{buttiker}T.~Christen, M.~B\"uttiker, to be published in
  Europhys.\ Lett.  
\bibitem{index}Considering a single channel we drop the index $\nu$.
\bibitem{note}The energy dependence of the Andreev reflection amplitude 
produces slight deviations close to the gap energy.
\bibitem{melsen}J.~A.~Melson, C.~W.~J.~Beenakker, Phys.\ B {\bf 203}, 219
  (1994).
\bibitem{bagwell}S. Chaudhuri and P.~F.\ Bagwell, Phys.\ Rev.\ B {\bf 51}, 16936
 (1995).
\bibitem{angle}This is connected to the decrease of the transmission 
probability 
with increasing incidence angle, and the non-isotropic distribution of the 
channels over the incidence angles\cite{bagwell}, which is $\propto 
\sin\left(2\vartheta\right)$.
\bibitem{gefen}J.~T.\ Edwards and D.~J.\ Thouless, J.\ Phys.\ C {\bf 5}, 807 
(1972); A.~Altland, Y.\ Gefen, G.\ Montambaux, Phys.\ Rev.\ Lett.\ {\bf 76},
 1130 (1996).
\bibitem{wees}B.~J.~Wees, P.~d.~Vries,
  P.~Magn\'ee, and T.~M.~Klapwijk, Phys.\ Rev.\ Lett.\ {\bf 69}, 510
  (1992).




 
\end{thebibliography}
\end{document}